\documentclass[%
 reprint,
superscriptaddress,
 amsmath,amssymb,
 aps,
]{revtex4-2}

\usepackage{graphicx}
\usepackage{dcolumn}
\usepackage{bm}

\begin{document}

\preprint{APS/123-QED}

\title{Merons Mediate Re-Ordering of Curved Rods Under Shear Flow}

\author{Nicholas W. Hackney}
\affiliation{Sandia National Laboratories, Albuquerque New Mexico 87185 USA}

\author{Joel T. Clemmer}
\affiliation{Sandia National Laboratories, Albuquerque New Mexico 87185 USA}

\author{Thomas O'Connor}
\affiliation{Department of Material Science and Engineering, Carnegie Mellon University, Pittsburgh, Pennsylvania, 15123}

\author{Gary S. Grest}
\affiliation{Sandia National Laboratories, Albuquerque New Mexico 87185 USA}%

\date{\today}

\begin{abstract}

Bent-core liquid crystals are a canonical example of a soft matter system whose behavior is controlled by a local preference for order that cannot be universally achieved. This geometric frustration, arising from the rod's curved shape, has been shown to stabilize a variety of equilibrium phases, such as the helically ordered nematic twist-bend phase ($N_{\rm TB}$). Unlike traditional nematics, the twist-bend state has 1D translational order arising from a periodic rotation of bend orientation along the helical axis. Here, we use molecular dynamics simulations to study the effect of shearing the $N_{\rm TB}$ phase along directions parallel and perpendicular to the helical axis. In the case of shear perpendicular to the helical axis, the nematic twist-bend phase is stable and flows without disordering. Conversely, shear along the helical axis disrupts order and leads to the emergence of fractionally charged Skyrmion defects, i.e. merons. These defects act as topological machines, locally rotating rods into a re-ordered and stable orientation of the $N_{\rm TB}$ phase. These findings reveal a new mechanism to create and control merons and highlight the potential application of bent-core liquid crystals in designing functional material with specific optical and computational properties.
\end{abstract}

\maketitle

\section{Introduction}\label{sec1}

Topological defects are a ubiquitous feature of condensed matter system and have been recognized to play an important role in mediating changes of order across thermodynamic phase transitions. For example, xy-vortices mediate the loss of quasi-long range order in the Berezinski-Kosterlitz-Thouless transition~\cite{kosterlitz1973ordering}, dislocation unbinding plays a critical part in two-dimensional melting~\cite{PhysRevB.19.2457} and string-like defects have been observed in the isotropic-nematic transition of unixial liquid crystals~\cite{bowick1994cosmological}. More recently, topological defects have been shown to perform the same function in states of driven~\cite{vyas2026two} or active flow~\cite{duclos2020topological}. These out-of-equilibrium systems have the added feature that defects can couple to flow in a non-trivial way, leading to behavior not seen in their equilibrium counterparts. For instance, dislocations experience a transverse Peach-Koehler force in response to external stress~\cite{peach1950forces}, positively charged point defects have enhanced motility in 2D active nematics~\cite{PhysRevLett.121.108002}, charge neutral disclination loops exhibit complex dynamics in 3D active nematics~\cite{vcopar2019topology,duclos2020topological}, and chiral defects exhibit the Hall effect in 2D ferro-magnets~\cite{schulz2012emergent,jiang2017direct}. While this coupling is well understood in the aforementioned examples, the emergence and behavior of chiral defects under 3D flow remains a frontier problem. To address this, we use molecular dynamics simulations to study colloidal bent-core liquid crystals as a model system that forms chiral defects while also exhibiting fluid-like behavior in response to shear.

In systems with orientational order described by a 3D director field $\textbf{n}$, chiral defects are manifest as doubly twisted textures with integer or fractional topological charge, $Q=\frac{1}{4\pi}\int \rm dA~\mathbf{n}\cdot\partial_x\mathbf{n}\times\partial_y\mathbf{n}$, i.e.~Skyrmions or merons~\cite{selinger2016introduction,duzgun2018comparing,binysh2020geometry}. These topologically stable textures were originally defined as a model to describe the stability of nuclear particles~\cite{skyrme1962unified} and have since been reified in several condensed matter systems possessing either a broken inversion symmetry or Dzyaloshinski-Moriya interactions~\cite{duzgun2018comparing}. For example, Skyrmionic defects have been identified with 2D spin textures in chiral ferromagnets~\cite{roessler2006spontaneous,muhlbauer2009skyrmion,yu2010real}, particle like excitations in Bose-Einstein condensates~\cite{al2001skyrmions,LesliePhysRevLett.103.250401} and ordered lattices of double twist cylinders in the so-called blue phases of cholesteric liquid crystals~\cite{fukuda2011quasi,nych2017spontaneous,pivsljar2022blue,wright1989crystalline}. Furthermore, these defects have attracted a great deal of attention for their  memory storage capability~\cite{fert2013skyrmions,fert2017magnetic,han2022high} and application in creating displays with fast optical switching~\cite{kikuchi2002polymer}.

\begin{figure*}[ht!]%
\includegraphics[width=\textwidth]{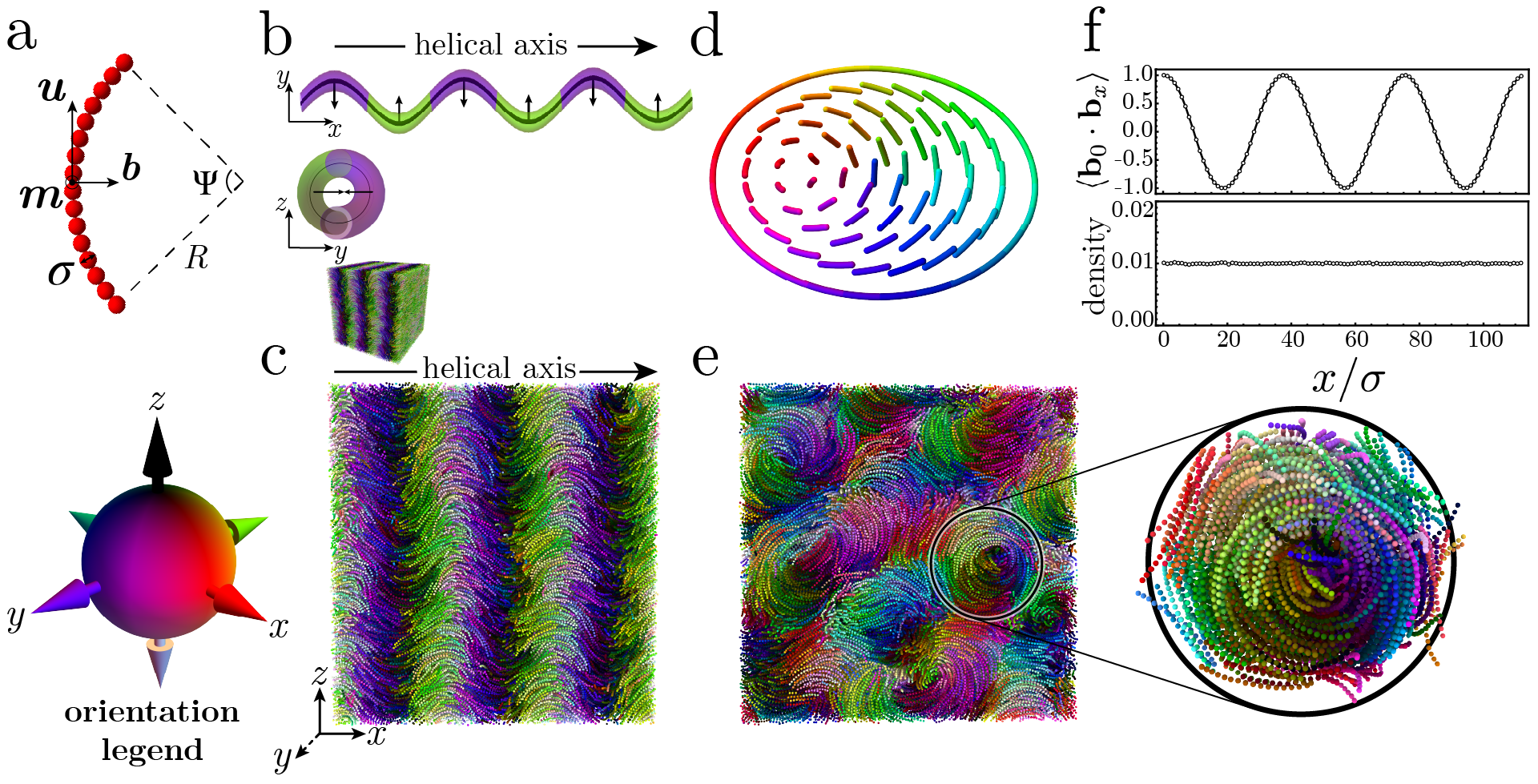}
\caption{\textbf{Schematic illustration of twist-bend textures observed in bent-core liquid crystals.} (\textbf{a}) Curved rods modeled as string of sixteen beads with diameter $\sigma$ tracing out a circular arc with radius $R$. Opening angle $\Psi$ defines the dimensionless ratio of rod length to radius of curvature. Three orthonormal vectors $\mathbf{u},\mathbf{b},\mathbf{m}$ needed to describe rod orientation illustrated on top of curved rod. (\textbf{b}) Helical integral curve of nematic twist-bend state. Green and purple segments represent approximate tiling of helices with curved rods. Arrows denote orientation of the bend director of each rod. (\textbf{c}) Bulk nematic twist-bend phase obtained from molecular dynamics simulations. (\textbf{d}) 2D slice of nematic director field of meron double twist cylinder illustrated with curved rods. At the center of the cylinder, rods point perpendicular to cylindrical slice and twist as they move radially outward until they are lying in plane at the edge of the cylinder. (\textbf{e}) Meta-stable amorphous blue phase obtained from molecular dynamics simulations. Inset highlights meron defect with local double twist texture. (\textbf{f}) Average bend correlation and mass density of simulated nematic twist bend phase highlighting that the colored bands observed in \textbf{c} represent periodic phase ordering rather than periodic density modulation characteristic of smectic ordering. Rods throughout this manuscript are colored by the direction the normalized bend director points to on the colored sphere illustrated in the orientation legend, with the hue denoting orientation in the $xy$ plane and the lightness denoting the $z$ component of the bend. Simulations shown in (\textbf{c}) and (\textbf{e}) performed with $N=40500$ rods and pressure $P=0.5\epsilon/\sigma^3$. Details related to the preparation of these two states provided in the Appendix~\ref{MD appendix}. } \label{fig: schematic illustration}
\end{figure*}

Recently, these chiral double twist defects have been shown to form in bent core liquid crystals composed of achiral rods with uniform curvature~\cite{subert2024achiral,HackneyD6SM00016A}. Here, the shape of the rods encourages them to arrange into configurations with uniform bend. However, such arrangements cannot smoothly fill two or three dimensional Euclidean space and must be coupled with additional splay or twist deformations~\cite{edwards1976molecular,dozov2001spontaneous,niv2018geometric}. This has been shown to give rise to a plethora of unique ordered phases, depending on the temperature, density and geometry of the individual rods (see Fig.~\ref{fig: schematic illustration}a)~\cite{takezoe2006bent,jakli2018physics,fernandez2024liquid}.

Of particular interest to the study of chiral structures is the specific case of curved colloidal rods, which exhibit a preference for twist-bend coupling under conditions of intermediate density and rod curvature~\cite{chiappini2019biaxial,chiappini2021generalized,kubala2022silico,subert2024achiral,HackneyD6SM00016A}. This coupling can manifest itself in the formation of a chiral nematic twist-bend ($N_{\rm TB}$) phase characterized by a director field that traces out helices of constant curvature and torsion. In bulk three dimensional systems, these helical integral curves are easily tiled by curved rods which can pack together to form a uniform ordered phase. A schematic illustration of this heliconical director field and the resultant nematic twist-bend phase is provided in Figure~\ref{fig: schematic illustration}b\&c. Here, the rods are colored by the direction their bend director, $\mathbf{b}$, points to when placed in the center of the hue, saturation and lightness sphere illustrated in the orientation legend of Figure~\ref{fig: schematic illustration}. This coloring scheme is used throughout the remainder of this work. Lastly, it is important to point out that, while the $N_{\rm TB}$ phase has little or no smectic order, translational symmetry is nonetheless broken by the periodic winding of the rod's bend orientation along the helical integral curves.

Alternatively, the coupling of twist-bend deformations can also result in the formation of chiral defects, which appear in the form of disordered line-like networks with a locally double twist texture, i.e. the amorphous blue phase (see Fig.~\ref{fig: schematic illustration}d\&e). These defects were found to be at least strongly metastable in bulk and occur over a range of density and rod curvature similar to that of the nematic twist-bend phase. Thus, it is natural to ask whether such double twist defects can emerge as a result of flow disrupted order in the nematic twist-bend phase. Furthermore, the topological charge of these meta-stable defect lines has yet to be characterized. However, the details of how they embed themselves within a heliconical background field depend on their charge~\cite{binysh2020geometry}. Therefore, an explicit characterization of the charge of any emergent defect would shed light on their role in mediating order transitions and help elucidate their dynamics under flow.

\begin{figure*}[ht!]%
\includegraphics[width=0.9\textwidth]{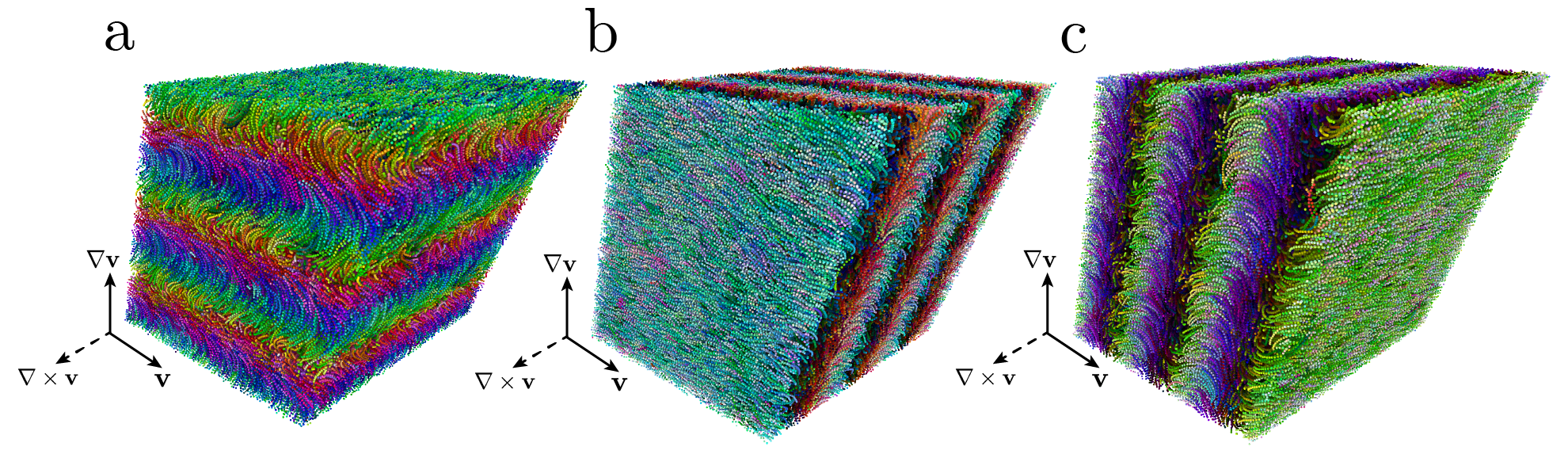}
\caption{\textbf{Orientation of nematic twist-bend phase with respect to shear.} Examples of the nematic twist-bend phase shown with the helical axis oriented along (\textbf{a}) the velocity gradient direction, (\textbf{b}) the vorticity direction and (\textbf{c}) the velocity flow direction. } \label{fig: orientations}
\end{figure*}

Here, we use molecular dynamics simulations to investigate the effects of shear flow on the nematic twist-bend phase of bent-core liquid crystals. Specifically, we consider shearing along three distinct directions with respect to the translational order. We find that the response to shear can be characterized according to the direction of flow with respect to the helical axis. In the two cases where shear is oriented perpendicular to the twist-bend helices, the nematic twist-bend phase is stable and the rods are able to flow without disordering. Interestingly, both of these orientations exhibit a chiral transport of the helical phase perpendicular to the direction of flow and independent of the motion of the rod's center of mass. Alternatively, when shear is oriented along the helical axis, we find that the nematic twist-bend phase is unstable. The system exhibits a finite yield stress peak and eventually re-orders into a steady state configuration corresponding to one of the stable orientations. This re-ordering process is marked by the emergence of fractionally charged meron defects with a chiral double twist texture. 

Following the formation and flow of these merons reveal that the rods are constantly diffusing in and out of the region near the meron core, indicating that the topological texture exists independently of the arrangement of any finite, fixed set of rods. Moreover, the rods that diffuse away from the core tend to assume their final nematic twist-bend ordering, suggesting that the merons serve as topological machines that locally rotate the twist-bend helices into their final orientation.

\section{Results}

In order to explore the effects of shear flow on the nematic twist-bend phase, we first generate ordered initial configurations. This is done by performing molecular dynamics simulations at constant rod number, pressure and temperature (see Appendix~\ref{MD appendix} for full details). The rods are modeled as approximately rigid strings of 16 beads with diameter $\sigma$. These beads trace out a circular arc with radius $R$ to form curved rods with opening angle $\Psi=L/R=1.5$. As the radius of curvature breaks the azimuthal symmetry of the rods, three orthonormal vectors $\mathbf{u}$, $\mathbf{b}$, and $\mathbf{m}$ are needed to describe their local orientation~\cite{xu2020general}. A schematic illustration of the rod geometry is provided in Figure~\ref{fig: schematic illustration}a. 

Previous studies of colloidal bent-core liquid crystals have established that the equilibrium phase behavior is controlled by the rod density $\phi$ and opening angle~\cite{memmer2002liquid,anzivino2020landau,kubala2022silico}. Rods with intermediate curvature (i.e. $\psi\in[1.2,1.9]$) exhibit a sequence of isotropic$\rightarrow$nematic twist-bend$\rightarrow$smectic X or splay-bend phases under increasing concentration. A contemporary study~\cite{HackneyD6SM00016A} of approximately rigid rods interacting via a purely repulsive Lennard-Jones (LJ) potential found that rods with the same geometry specified here assemble into the nematic twist-bend phase over a density range of  $\phi\in[0.42,0.56]$. Thus, a nematic twist-bend phase with $N=40500$ rods and density $\phi\simeq0.51$ is created via the initialization scheme introduced in Ref.~\cite{HackneyD6SM00016A} and outlined in Appendix~\ref{MD appendix}.

A visual example of the nematic twist-bend phase is provided in Figure~\ref{fig: schematic illustration}c. Visualizing the simulations in this way immediately reveals a one-dimensional translational ordering of rod orientation manifested as a periodic stripe pattern. The average bend correlation and mass density along the helical axis (see Fig.~\ref{fig: schematic illustration}f) confirms that these patterns are the result of periodic ordering of bend orientation and not a smectic ordering of rod density.

The effects of shear are examined by fixing the volume and deforming the simulation box in the xz-plane at a constant strain rate of $\dot{\gamma}=10^{-4}\tau^{-1}$. This establishes a velocity flow field of $\mathbf{v}=\dot{\gamma}z\hat{x}$, which has gradients along the $\hat{z}$-direction and non-zero vorticity oriented along the $\hat{y}$-direction. The rods are given an initial velocity that matches the imposed flow field. This is done to prevent non-physical boundary effects that can result from an improper remapping of velocity across the top and bottom of the simulation box when shear is suddenly turned on. Also, in order to avoid the system becoming too skewed, the triclinic box is confined to xz tilt angles between $-45^\circ$ and $45^\circ$. As configurations with these two tilt factors are geometrically equivalent, the simulation box can be remapped between these two angles, allowing simulations to run to arbitrarily large strain~\cite{kraynik1992extensional}. 

With the direction of shear fixed, the nematic twist-bend phase can be rotated to explore the effects of shearing along different directions with respect to the periodic phase ordering. To that end, we consider three different orientations of the $N_{\rm TB}$ phase, defined by the alignment of the helical axis along: a) the velocity gradient direction, b) the vorticity direction and c) the velocity flow direction. A schematic illustration of each orientation is provided in Figure~\ref{fig: orientations}. 

Using each of these orientations as a starting state, shear simulations are carried out to a strain of $\gamma\sim10$. The response of each orientation is characterized by measuring the off diagonal components of the stress tensor, $\mathbf{\sigma}$. In addition, the liquid crystalline order is monitored by measuring the average nematic $\langle S\rangle$, helical $\langle\vert\tau\vert\rangle$ and smectic $\langle \rho \rangle$ order parameters (see Appendix~\ref{op appendix} for definition and discussion). 

The three orientations discussed here can be grouped according to whether or not they have translational order along the shear direction. As translational order plays an important role in governing how a material responds to shear, the rest of the discussion is broken up around this grouping.

\begin{figure*}[ht!]%
\includegraphics[width=\textwidth]{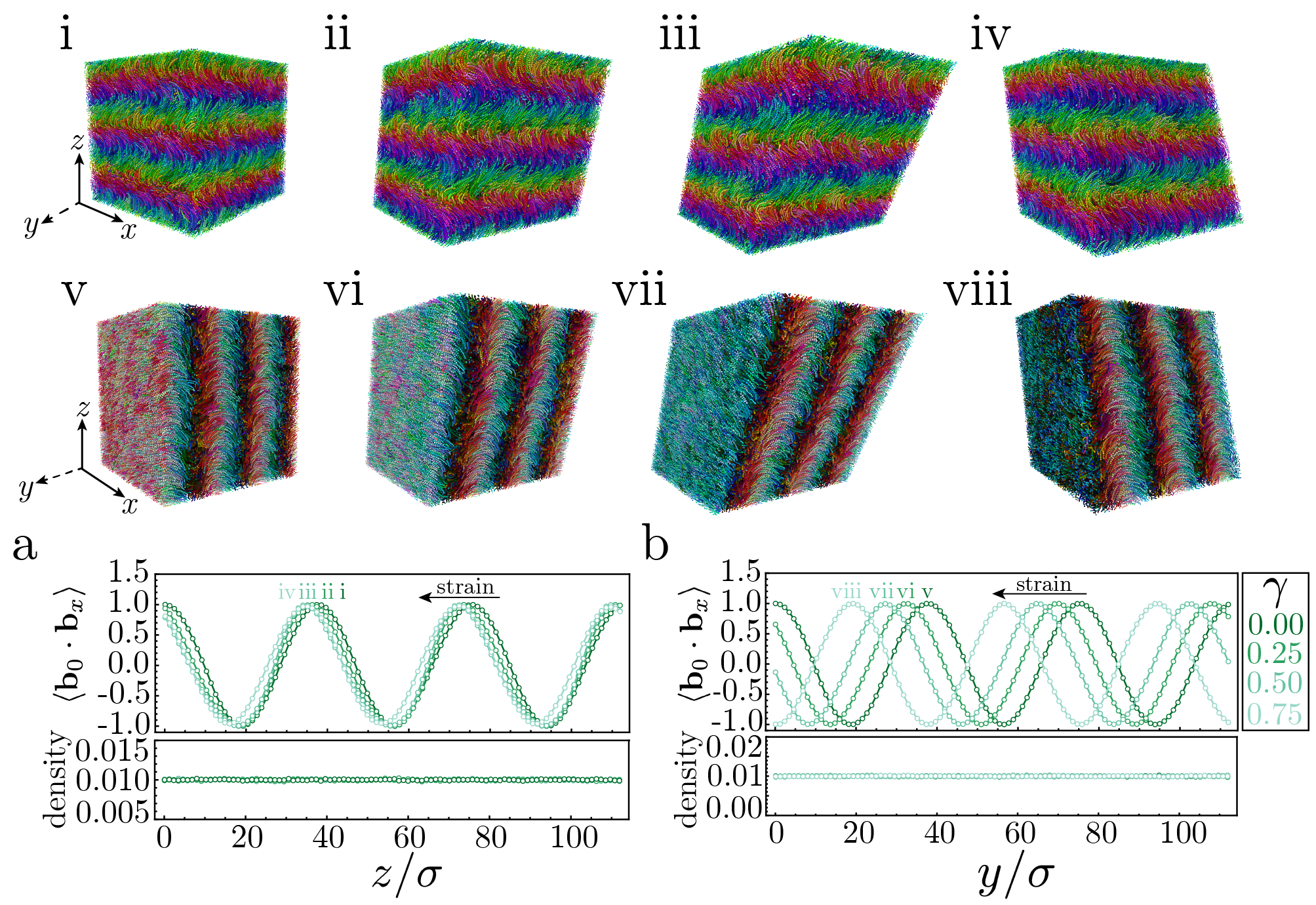}
\caption{\textbf{Shear perpendicular to the helical axis of the nematic twist-bend phase.} Visualizations of the sheared $N_{\rm TB}$ phase with periodic phase direction oriented along the velocity gradient direction (i-iv) and the vorticity direction (v-viii). Videos of the full simulations provided in Supplementary Movies 1\&2. The average bend correlation and mass density along the velocity gradient (\textbf{a}) and the vorticity (\textbf{b}) direction are shown for several different values of strain. All simulations run with $N=40500$ rods and strain rate $\dot{\gamma}=10^{-4}\tau^{-1}.$ } \label{fig: fluid like orientations}
\end{figure*}

\subsection{Shear Perpendicular to the Helical Axis}

Translational ordering along the shear direction is absent in the two cases where the helical axis is oriented along the velocity gradient and vorticity direction. As a result, one would not expect either of these to exhibit an elastic response to shear. Indeed, in both cases, the only non-zero off-diagonal component of the stress tensor ($\sigma_{xz}$) immediately plateaus to a small value of stress (see Fig.~\ref{SI: fluid OP}a\&c in Appendix~\ref{op appendix}). In addition, preservation of the translational order in these two orientations suggests that the nematic twist-bend state may not be broken up at all. This is confirmed by looking at the average nematic, helical and smectic order which remain constant across the entire range of strain (see Fig.~\ref{SI: fluid OP}b\&d in Appendix~\ref{op appendix}). In Figure~\ref{fig: fluid like orientations}i-iv and Supplementary Movie 1, we show simulation visuals for the case where the helical axis is aligned with the velocity gradient direction. Similarly, Figure~\ref{fig: fluid like orientations}v-viii and Supplementary Movie 2 show the same sequence for the case where the helical axis aligns with the vorticity direction. These visuals confirm that the nematic twist-bend phase is stable under shear for both orientations.

While this sequence of strain images ostensibly suggests that nothing happens to either of these orientations under shear, a close inspection reveals a drift in the colored bands along the helical axis. For example, the dark red bands in Fig.~\ref{fig: fluid like orientations}v-viii move to the left with increasing strain. As the colors are related to the orientation of the rod's bend director, this suggests that this drift is the result of a coherent rotation of rods and not a uniform mass transport. In Figure~\ref{fig: fluid like orientations}a\&b we show the bend correlation and rod density as a function of strain. The waves in bend orientation travel to the left with increasing strain. Conversely, we observe no strain dependent change in rod density. This signifies that the apparent layer drift corresponds to a corkscrew-like transport of helical phase perpendicular to the flow direction. The strain dependence of this phase transport is noticeably faster when the helical axis is oriented along the vorticity direction than when it is oriented along the velocity gradient direction.

\begin{figure*}[ht!]%
\includegraphics[width=\textwidth]{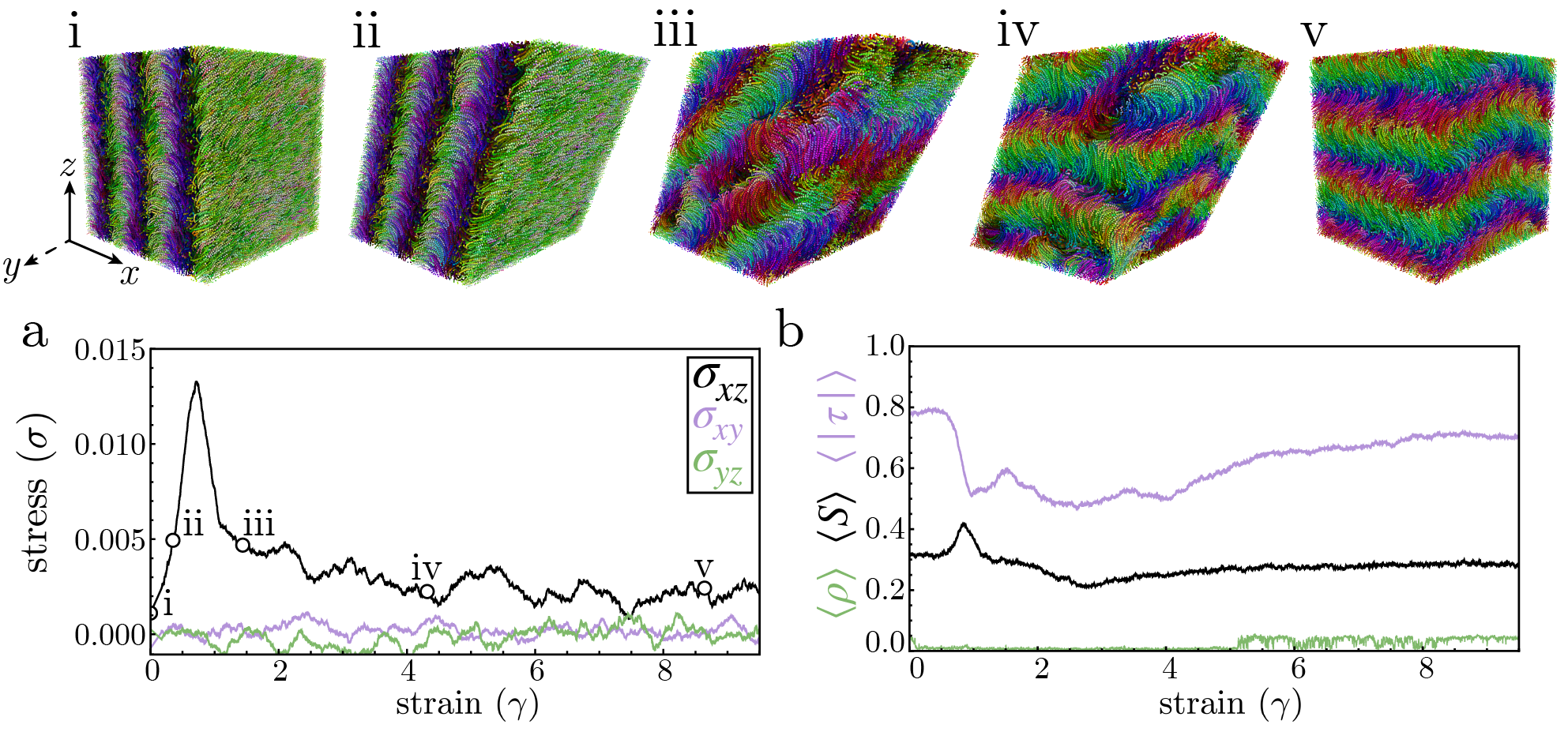}
\caption{\textbf{Shear parallel to the helical axis of the nematic twist-bend phase.} Simulation snapshots of sheared $N_{\rm TB}$ phase with periodic phase direction oriented along the velocity flow direction (i-v). Videos of full simulations provided in Supplementary Movie 3. (\textbf{a}) Off diagonal elements of the stress tensor shown as a function of strain. (\textbf{b}) Average helical $\langle\vert\tau\vert\rangle$, nematic $\langle S \rangle$ and smectic $\langle\rho\rangle$ order parameter as a function of strain. All simulations run with $N=40500$ rods and strain rate $\dot{\gamma}=10^{-4}\tau^{-1}.$ } \label{fig: solid like orientation}
\end{figure*}

\subsection{Shear Parallel to Helical Axis}

In contrast to the two previously considered orientations, shear parallel to the helical axis disrupts nematic twist-bend order. Simulation images for this orientation are shown in Fig.~\ref{fig: solid like orientation}i-v for a representative sequence of increasing strain. In addition, full simulation videos are provided in Supplemental Movie 3. At low strains, we observe a uniform tilting of the phase pseudo-layers. Eventually, these layers appear to merge and break up but never fully disappear. At intermediate strain, the dislocated layers still exist but appear to be rotated. Finally, in the limit of large strain (i.e. $\gamma\sim10$), the pseudo-layer dislocations disappear. The system is once again in an ordered nematic twist-bend phase, however, the helical axis has been rotated such that it aligns parallel to the velocity gradient direction. 

The off-diagonal components of the stress tensor for this orientation are plotted as a function of strain in Fig.~\ref{fig: solid like orientation}a. Similar to the other orientations, $\sigma_{xz}$ is the only non-zero off-diagonal component. The system exhibits a yield stress peak in this component of the stress tensor. This peak is indicative of a regime of elastic deformation of the nematic twist-bend order that persists up to a strain of about $\gamma\simeq0.7$. At the yield stress peak, the system begins to suffer irreversible plastic deformations, coinciding with the apparent merging of phase pseudo-layers. Past this, the stress drops to some small value which roughly corresponds to the average stress measured for the case where the helical axis points along the velocity gradient direction. As the behavior observed here is markedly different than that reported for the other two orientations, we performed three separate trials of shearing along the helical axis. All three of which displayed the same behavior and the results are provided in Figure~\ref{SI: stress vs strain trials} located in Appendix~\ref{MD appendix}.

In Figure \ref{fig: solid like orientation}b, we show the average nematic, helical and smectic order parameters as a function of strain. Similar to the conclusions drawn from the simulation visuals, we observe little to no change in the order up to the yield stress peak. Then, at the onset of plastic deformation, we observe a sudden loss in helical order that is accompanied by a brief increase in nematic order. Beyond the yield stress peak, both continue to decrease slightly. However, the system never becomes fully disordered and the nematic and helical order gradually rise back to around their initial value as the system returns to the nematic twist bend phase. While the nematic twist-bend phase has very little smectic order to begin with, a sudden drop and subsequent rise in smectic order is observed as the initial twist-bend phase is disrupted and eventually reforms.

The disruption of the phase pseudo-layers across the yield stress peak make it clear that the translational order of the rod's bend orientation is broken up by shear parallel to the periodic phase ordering. In strained smectic liquid crystals, this loss of order coincides with the emergence of dislocations in the translational ordering of particle position. While the nematic twist-bend phase lacks positional order, the periodicity of the helical phase has been shown to have the same elastic energy as a smectic on scales much larger than the helical pitch~\cite{kamien1996liquids,ParsouziPhysRevX.6.021041,meyer2016local}. As a result, the $N_{\rm TB}$ phase can support all the same defects as a smectic liquid crystal. This suggests that the disruption and eventual re-emergence of the nematic twist bend phase can also be understood as a defect driven process.

\begin{figure*}[ht!]%
\includegraphics[width=\textwidth]{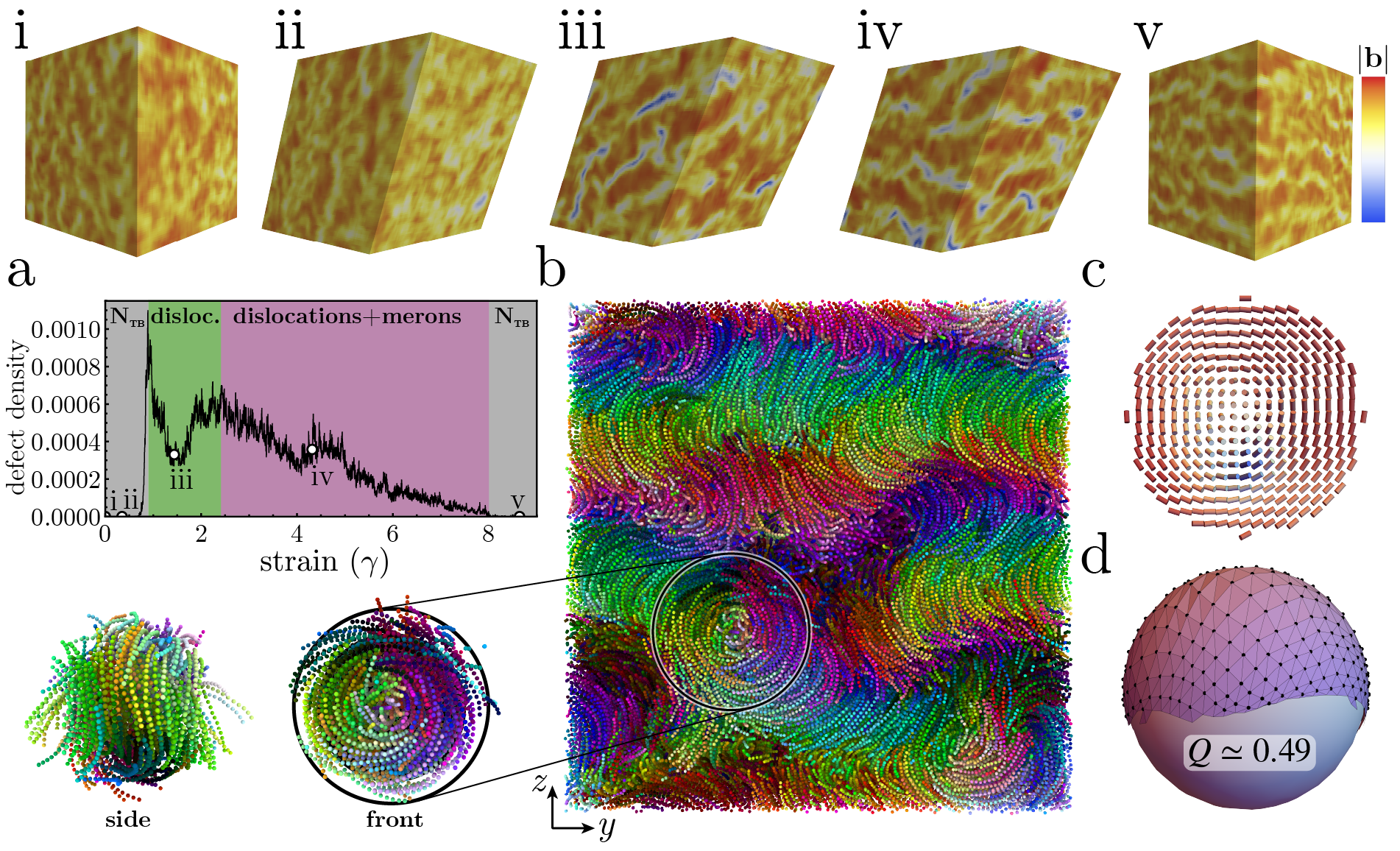}
\caption{\textbf{Defects in sheared nematic twist-bend phase.} (\textbf{a}) Defect density as a function of strain. Colored background denotes range over which each important feature is observed. Labeled points $i-v$ correspond to sequence of strained states represented as coarse grained lattice of rod bend director. Lattice sites colored by the magnitude of the coarse grained bend, $\vert\mathbf{b}\vert$. These states correspond to the same sequence provided in Fig. \ref{fig: solid like orientation}i-v. Videos provided in Supplemental Movie 4. (\textbf{b}) Image of xy-plane slice from sheared twist-bend simulation with strain of $\gamma\simeq4.8$. Inset highlights defect with local meron texture. (\textbf{c}) Coarse grained nematic director of isolated defect structure shown in inset of ($\textbf{b}$). Here, the cylinders are colored according to the magnitude of bend. (\textbf{d}) Gaussian map of nematic director showing Skyrmion charge represented as surface area fraction of unit sphere tiled by triangular mesh defined by the nematic director.} \label{fig: defects}
\end{figure*}

\subsection{Topological Defect Analysis}

In order to identify smectic and meron-like defects, we coarse grain the simulation data to obtain a lattice of average bend and nematic directors (see Appendix~\ref{cg appendix}). We identify line-like geometric degeneracies in the rod bend orientation as connected clusters of lattice sites with vanishing bend. In liquid crystalline systems with local preference for bend, these line-like structures are associated with a variety of topological defects such as screw/edge dislocations, Skyrmions, merons and Hopfions~\cite{binysh2020geometry}. As a result, identification of these bend zeros allows for the characterization of these intricate 3D textures in terms of simplified 1D structures. Furthermore, the emergence of defects in the $N_{\rm TB}$ phase is easily quantified by measuring the fraction of defective sites in the coarse grained lattice. 

In Figure ~\ref{fig: defects}a, this defect density is plotted as a function of strain for the case of the nematic twist-bend phase with the helical axis oriented along the velocity flow direction. From this, it is clear that the system flows defect free during the initial elastic deformation. Then, the yield stress peak induces bend instabilities which nucleate topological defects. At intermediate strain, we observe that the defect density has several subsequent peaks before gradually dropping to zero as the system re-orders. Figure \ref{fig: defects}i-v shows several coarse grained lattices of bend magnitude for the same sequence of states in Figure \ref{fig: solid like orientation}i-v. The defects nucleated at the yield stress peak are manifested as long lines of degenerate (i.e. $\vert\mathbf{b}\vert\sim0$) bend. These defects merge and rotate along the direction of flow before eventually disappearing.

To distinguish between bend degeneracies related to merons and smectic-like dislocations, we measure the topological charge around each defect line. As the analysis required to measure the charge of Skyrmionic defects from molecular dynamics simulations is\textemdash to our knowledge\textemdash new and constitutes a central contribution of this work, an extended discussion of the relevant details is included in Appendix~\ref{charge appendix}. Using this, we find that all of the observed defect lines have zero topological charge up until the second peak in defect density at $\gamma\simeq2.5$. Taken with the observation that this initial period of yielding is characterized by the recombination of helical pseudo-layers, the lack of charge confirms that the defects observed until this point are screw/edge dislocations in the nematic twist-bend phase. At the second peak in defect density, we observe the emergence of a small number of defect lines with half integer charge that persist alongside the other defects until the defect density drops to zero. In Figure \ref{fig: defects}b, we show the cross-section of a simulation containing a defect line with fractional topological charge. Notably, this cross-section contains a vortex-like structure\textemdash similar to that found in the amorphous blue phase shown in Figure \ref{fig: schematic illustration}e\textemdash which coincides with the location of the bend zero. Figure~\ref{fig: defects}c shows the coarse-grained nematic director on a disc around this bend zero. This reveals a distinctive double twist profile similar to that illustrated in Figure \ref{fig: schematic illustration}d. 

Conceptually, the Skyrmion charge measures the fraction of the unit sphere covered by the Gaussian map of $\mathbf{n}$ over a 2D measuring surface. Thus, the Gaussian map of a $Q=1$ Skyrmion covers the entire sphere while that of a $Q=1/2$ meron only covers half~\cite{selinger2016introduction}. Indeed, looking at the triangulation of the Gaussian map in Figure ~\ref{fig: defects}d reveals a near perfect hemispherical cap, confirming the classification of this defect as a meron. Interestingly, all of the defect lines that we observe with non-zero topological charge are categorized as merons, suggesting that disrupting the nematic twist-bend phase via shear flow selectively nucleates merons instead of Skyrmions.

\begin{figure*}[ht!]%
\includegraphics[width=\textwidth]{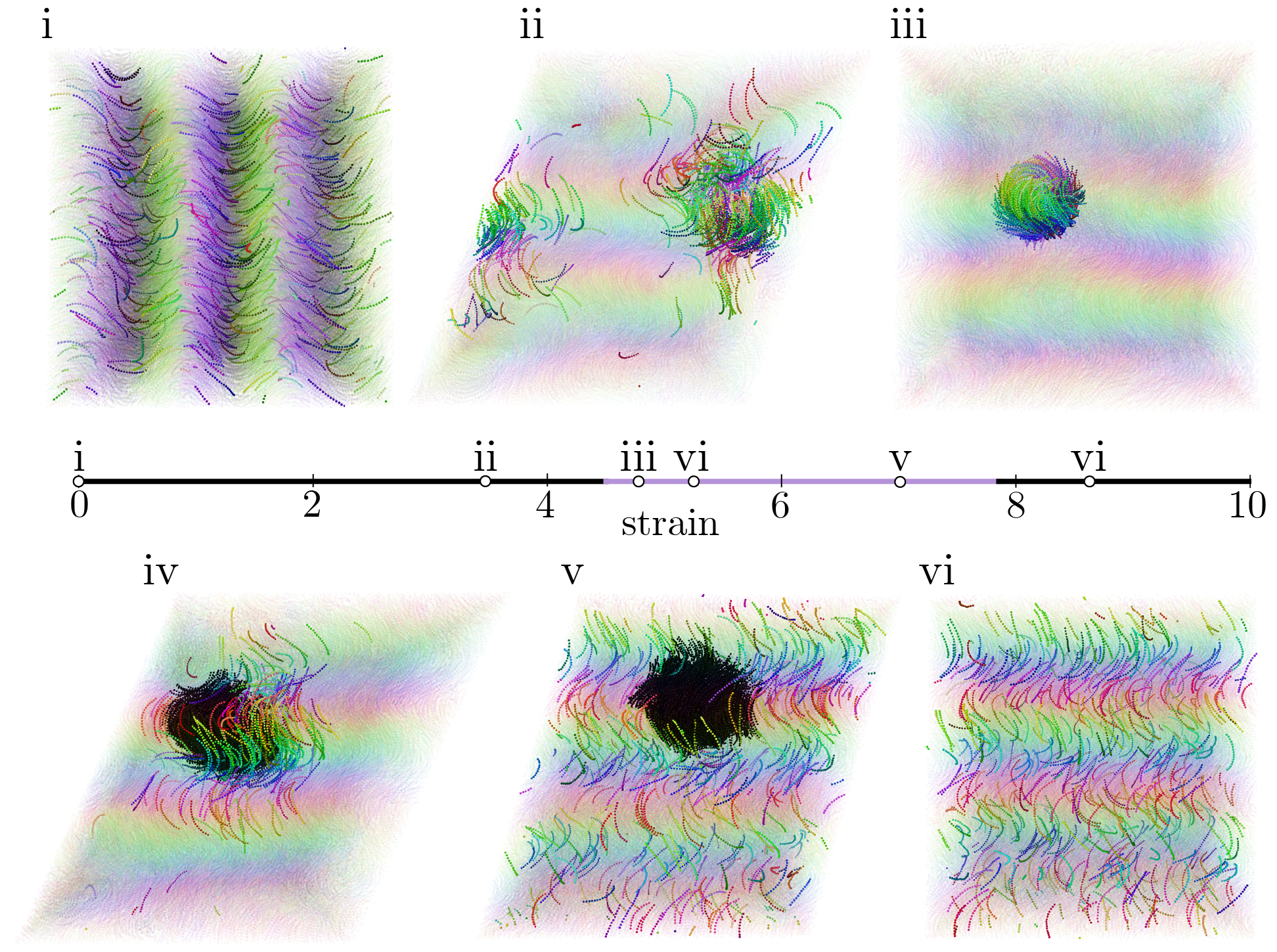}
\caption{\textbf{Meron formation and flow under shear.} Strain scale for shear simulation with strain range over which a particular meron is observed denoted by the purple line segment. Sequence of states highlighting just the rods that initially form into the meron shown in i-vi. Rods that make up meron after initial formation drawn in black to highlight that the topological texture persists despite rods constantly diffusing in and out of the defect. All other rods not involved in the meron are made translucent. Note that images only highlight a single meron and other defects may be present throughout the simulation shown here. Full video provided in Supplementary Movie 5.} \label{fig: merons}
\end{figure*}

\subsection{Meron Behavior Under Shear Flow}

To better understand how merons behave under shear flow and uncover the role they play in rotating the nematic twist-bend phase, we follow the set of rods that form into a meron at a strain $\gamma\simeq4.5$. In Figure~\ref{fig: merons}i-vi and Supplementary Movie 5, these rods are highlighted in a sequence of simulated states showing the formation, flow and eventual break up of a particular meron.  Initially, the rods are distributed over the entire system and slowly converge until they assemble into a tight, doubly twisted group around the defect core. Immediately after forming, the original set of rods begin to diffuse away from the defect core. However, the meron does not break up and the defect persists as rods continue to diffuse in and out of the region around the core. To emphasize this, the rods that lie in this region but were not a part of the original set of rods are highlighted in black. In the context of a heliconical background field, defect lines with a locally meronic texture are edge dislocations between two differently oriented heliconical layers~\cite{binysh2020geometry}. Thus, as rods can easily diffuse in and out of this double twist texture, we see that merons act as topological machines that locally translate rods between two nematic twist-bend orientations. After forming, the meron continues to flow for a while before disappearing at $\gamma\simeq8$. Surprisingly, the meron does not appear to annihilate with another defect. While the meron's disappearance likely has something to do with one of the nematic twist-bend orientations running out of rods and thereby annealing the edge dislocation, the exact topological mechanism for this remains unknown. It is important to point out that while the meron shown here is neither the first nor only meron to form throughout the simulation, we chose to highlight a single defect for the sake of clarity.

\section{Discussion}\label{sec1}

 In this study, we demonstrated that chiral meron defects emerge as a result of disrupted helical order in the nematic twist-bend phase of bent-core liquid crystals. Specifically, we performed molecular dynamics simulations investigating the behavior of the nematic twist-bend phase under simple shear. As this phase has one-dimensional translational ordering of rod bend orientation (i.e. phase), we considered three different orientations of the initial state with respect to shear. The two orientations that lack translational order along the shear direction are stable under flow. These two orientations exhibit a transverse transport of rod phase, resulting from a corkscrew-like tumbling of twist-bend helices. The nematic twist-bend phase is unstable to shear oriented along the direction of translational order and initially breaks up before re-ordering into one of the stable orientations. This re-ordering occurs via a four stage process:
i) at low strain, the system undergoes elastic deformation of helical pseudo-layers; ii) at the onset of yielding, strain induced bend instabilities nucleate dislocations in translational phase order; iii) these dislocations orient along the flow direction and combine to form chiral defects with locally double twist texture; iv) eventually, these defects disappear leaving an ordered nematic twist-bend phase with helical axis aligned along the velocity gradient direction. Measurement of the discrete topological charge over the region around these double twisted defects confirms their classification as fractionally charged merons rather than Skyrmions. This distinction is notable, as merons are known to act as edge dislocations between two differently oriented regions of twist-bend order~\cite{binysh2020geometry}. Thus, taken with the observation that rods freely diffuse in and out of the region around the core, this suggests that merons act like topological machines that locally translate rods between two differently oriented regions of nematic twist-bend order.

While these results correspond to a specific strain rate, simulations performed at a faster rate of $\dot{\gamma}=10^{-3}$ establish an upper bound on the behavior reported here. In this case, we observe a similar yielding behavior when shearing along the helical axis, however the system never re-orders and remains in a turbulent meron state (see Supplemental Movie 6). Similarly, shearing the nematic twist-bend phase when the helical axis is aligned with the velocity gradient direction results in undulations in the the helical pseudo-layers; eventually leading to their breakup (see Supplemental Movie 7). This is likely the result of a Helfrich-Hurault instability~\cite{blanc2023helfrich}. In fact, additional evidence of this type of instability is provided by the slight bending of the helical pseudo-layers observed in the rotated $N_{\rm TB}$ phase (see Fig.~\ref{fig: solid like orientation}v). However, in this case, the instability is most likely the result of a geometric strain arising from a slight anisotropy of the fixed box size which is commensurate with the helical pitch in the original orientation but not the final.

There remain several open questions related to the specific details about the creation, annihilation, and interaction between merons. For example, it is clear that merons arise after the formation of a background network of smectic-like defects. However, there are several possible mechanisms through which they could form. Merons could emerge as a result of a shear induced recombination of one or more of these defect lines. Alternatively, they could form through a process similar to that recently found in chiral nematics, where merons are emitted by line-like geometric singularities in cholesteric pitch as they change the sign of their winding~\cite{lbbj-txnx,pollard2025defect}. Similarly, it is unclear how exactly these merons disappear, as they do not obviously appear to annihilate with other merons. The answer to both of these questions will depend on the conservation of topological invariants, which can be expressed through the Gauss-Bonnet-Chern theorem relating the charge to the oriented intersection number of line-like singularities with the measuring surface and the integrated torsion around the boundary~\cite{machon2016umbilic,binysh2020geometry}. This in turn will be intimately related to the specific topology of the meron core, which has been shown to support a wide array of different structures such as lines, loops, knots, and links in both bend-nematic~\cite{binysh2020geometry} and chiral liquid crystals~\cite{ackerman2017diversity,tai2019three,hall2026fusion}.

Currently, the topology of the meron cores observed here remains unknown. In fact, it is unclear if there is even a unique preference or if the merons exhibit a wide range of different core structures. While disentangling these questions could lead to better control and design of specific defect textures in bent-core liquid crystals, we leave the complete characterization of topological structure and distribution of these shear induced defects to future works.

Finally, the new method introduced here for creating merons via shear is particularly well suited for studying the dynamics of these defects under varying flow conditions. For example, Skyrmions are known to exhibit the quantum Hall effect in flowing 2D chiral ferromagnets~\cite{neubauer2009topological,schulz2012emergent}. Since this effect is the result of the specific topology of the spin-texture, it is reasonable to expect a similar phenomenon to occur in flowing bent-core liquid crystalline systems. Furthermore\textemdash as merons have been observed as ordered lattice of ``flux-tubes" in quasi-2D slabs~\cite{fernandez2021hierarchical,subert2024achiral}, disordered networks in the bulk amorphous blue phase~\cite{subert2024achiral,HackneyD6SM00016A} and as the localized structures observed here\textemdash this situation provides a particularly exciting opportunity to uncover the role that different core structures have on transport phenomena. Overall, we hope that these findings inspire future work using liquid crystalline systems to study the emergent structure and dynamics of this diverse class of topological defects and aid in the design of new functional material with advanced optical, transport and computational properties.

\section*{Conflicts of interest}
There are no conflicts to declare.

\section*{Data availability}
The data from the molecular dynamics simulations and associated analysis are available from the authors upon reasonable request.

\section*{Acknowledgements}

This work was performed in part at the Center for Integrated Nanotechnologies, an Office of Science User Facility operated for the U.S. Department of Energy (DOE) Office of Science. Sandia National Laboratories is a multimission laboratory managed and operated by National Technology \& Engineering Solutions of Sandia, LLC, a wholly owned subsidiary of Honeywell International, Inc., for the U.S. DOE’s National Nuclear Security Administration under contract DE-NA-0003525. The views expressed in the article do not necessarily represent the views of the U.S. DOE or the United States Government.

\appendix

\section{Molecular Dynamics Simulations}\label{MD appendix}

Curved colloidal rods are modeled as a string of $16$ beads with mass $m$. The beads trace out a circular arc with length $15\sigma$ and radius $10\sigma$. Here, $\sigma$ defines the unit of length. The geometry of the rods can be expressed in a meaningful way by defining the dimensionless opening angle $\Psi=L/R=1.5$. In order to avoid known issues with shearing elongated rigid rods~\cite{berry2021lees}, we model the interaction between neighboring beads on the same rod via a bonded particle model~\cite{wang2009new,clemmer2024soft} that models each bead as a finite sized sphere with internal orientational degrees of freedom. This allows the rods to have some elasticity that is parameterized by a normal, shear, twist, and bend stiffness respectively defined as ${k_{\rm s},k_{\rm s},k_{\rm t},k_{\rm b}}$. Throughout this work, we set $k_{\rm r}=1000\epsilon/\sigma^2$, $k_{\rm s}=500\epsilon/\sigma^2$, $k_{\rm t}=1000\sigma$ and $k_{\rm b}=1000\sigma$. These values were chosen as they were shown in a previous work to approximate the equilibrium behavior of perfectly rigid bent-core liquid crystals~\cite{HackneyD6SM00016A}.

Beads on different rods interact through a repulsive Lennard-Jones potential,
\begin{equation}
U=\begin{cases}
4\epsilon\big[\big(\frac{\sigma}{r}\big)^{12}-\big(\frac{\sigma}{r}\big)^6 \big] & r<r_{\rm c} \\
0 & r>r_{\rm c}
\end{cases}
\end{equation}
\noindent where $\epsilon$ defines the interaction energy and $r_{\rm c}=2^{1/6}\sigma$ defines the cutoff distance. This value of $r_{\rm c}$ was chosen such that interactions between rods are purely repulsive. Using this, molecular dynamics simulations were carried out using the LAMMPS software package~\cite{thompson2022lammps}. The equations of motion are integrated via the velocity-Verlet algorithm with timestep $\delta t=0.0035\tau_{\rm LJ}$, where $\tau_{\rm LJ}=(m\sigma^2/\epsilon)^{1/2}$ defines the standard time unit for a Lennard-Jones system. Additionally, the bead's internal orientational degrees of freedom are integrated with a Richardson iterator included in the LAMMPS bonded particle model. All simulations were run with periodic boundary conditions.

Following the initialization procedure developed in Ref.~\cite{HackneyD6SM00016A}, we compress an isotropic gas of $N=1500$ rods to pressure $P=1.0\epsilon/\sigma^3$ and temperature $T=1.0\epsilon/k_{\rm B}$. This results in the formation of a high density smectic phase, which we replicate to create a larger configuration with $N=40500$ rods. A nematic twist-bend phase is subsequently obtained by allowing the system to equilibrate at $P=0.5\epsilon/\sigma^3$. A Langevin thermostat~\cite{schneider1978molecular} with damping $1\tau_{\rm LJ}$ and Berendsen barostat~\cite{berendsen1984molecular} with damping $200\tau_{\rm LJ}$ are used to maintain the temperature and pressure. During equilibration, the simulation box is allowed to dilate and compress anisotropically. This is done to avoid any potential effects arising from incommensurability between the helical pitch and box dimension.

\begin{figure}[ht!]%
\includegraphics[width=0.47\textwidth]{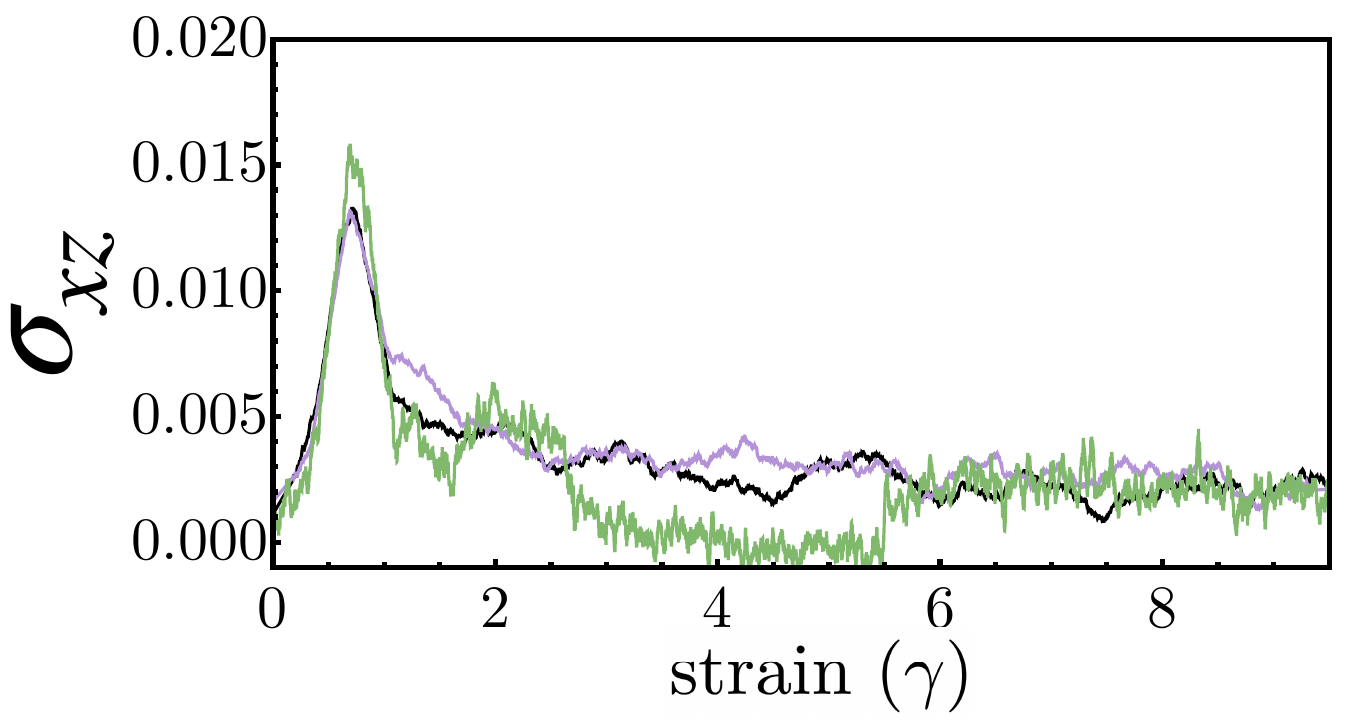}
\caption{ The $\sigma_{xz}$ component of the stress tensor plotted as a function of strain for three independent simulations of shearing the nematic twist-bend phase parallel to the helical axis. } \label{SI: stress vs strain trials}
\end{figure}

Once the initial state is obtained, the box size is fixed and can be rotated to obtain different orientations of the helical axis with respect to the x- y- and z-axis. The simulation box is then deformed in the xz-plane at constant engineering strain rate of $\dot\gamma=10^{-4}/\tau_{\rm LJ}$. Shear is enforced by shifting the particle velocity in the direction of flow as they move across the periodic boundary in the gradient direction.  The rods are given an initial velocity matching the shear profile to avoid non-physical boundary effects that arise from improper velocity remapping at the onset of shear. Simulations are run to a final strain of $\gamma\simeq10$. The off diagonal components of the stress tensor, $\sigma$, are measured for each shear simulation. Figure~\ref{SI: stress vs strain trials} shows the stress versus strain measured from three independent trials of shear along the helical direction of the nematic twist-bend phase.

\begin{figure*}[ht!]%
\includegraphics[width=\textwidth]{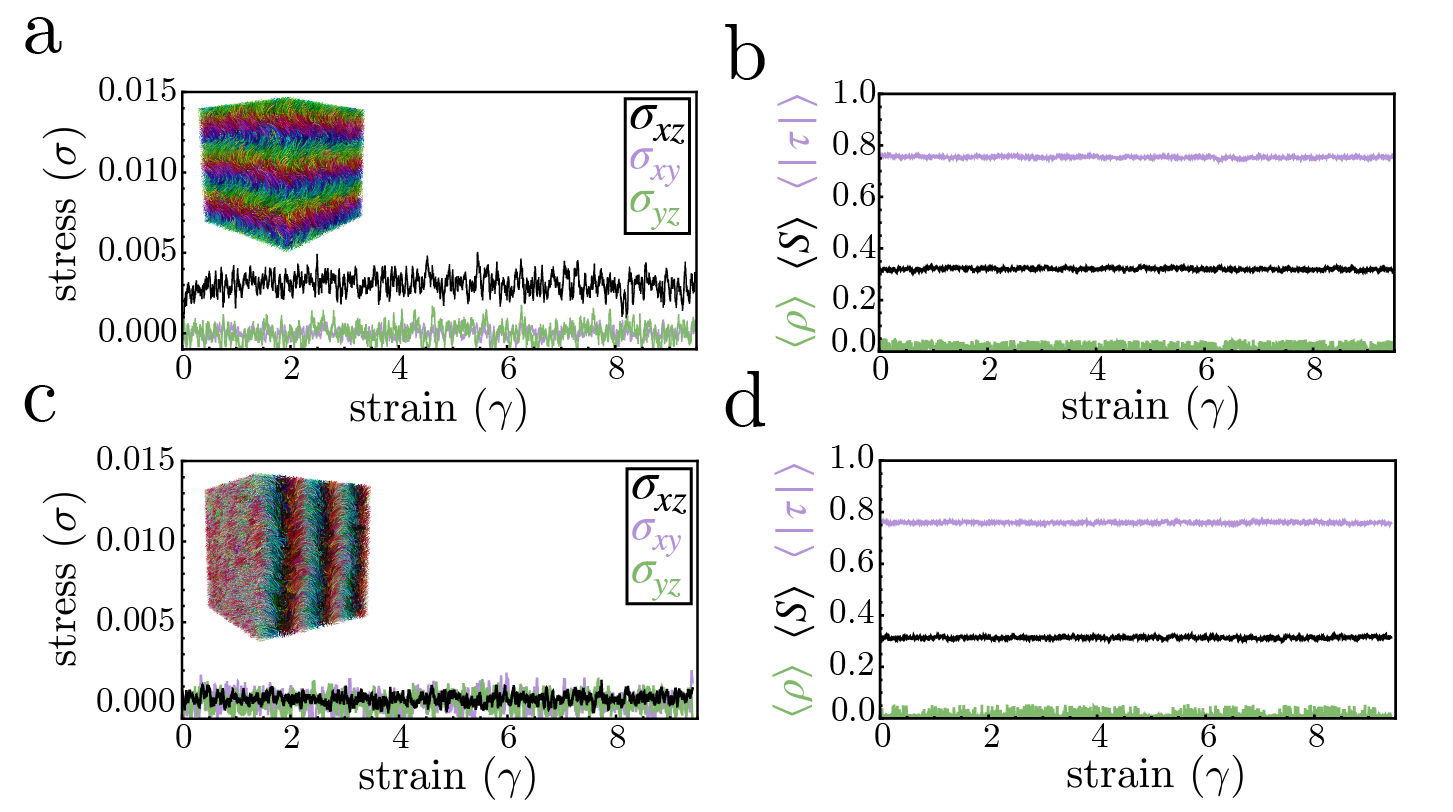}
\caption{\textbf{Shear parallel to the helical axis of the nematic twist-bend phase.}  (\textbf{a}) Off diagonal elements of the stress tensor shown as a function of strain for the orientation where the helical axis is aligned parallel to the velocity gradient direction. Inset shows initial state. (\textbf{b}) Average helical $\langle\vert\tau\vert\rangle$, nematic $\langle S \rangle$ and smectic $\langle\rho\rangle$ order parameter as a function of strain for the orientation where the helical axis is aligned with velocity gradient direction. (\textbf{c}) Off diagonal elements of the stress tensor shown as a function of strain for the orientation where the helical axis is aligned parallel to the vorticity direction. Inset shows initial state. (\textbf{d}) Average helical $\langle\vert\tau\vert\rangle$, nematic $\langle S \rangle$ and smectic $\langle\rho\rangle$ order parameter as a function of strain for the orientation where the helical axis is aligned with the vorticity direction. All simulations run with $N=40500$ rods and strain rate $\dot{\gamma}=10^{-4}\tau^{-1}.$} \label{SI: fluid OP}
\end{figure*}

\section{Order Parameters}\label{op appendix}

The liquid crystalline order is determined by measuring the average nematic $\langle S \rangle$, helical $\langle \vert\tau\vert \rangle$ and smectic $\langle \rho \rangle$ order parameters. Here, $\langle \cdots\rangle$ denotes the ensemble average of every rod in the system. The nematic order parameter is defined as the largest eigenvalue of the $\mathbb{Q}$-tensor,
\begin{equation}
\mathbb{Q}=\frac{1}{N}\sum_{i}^N\bigg(\frac{3}{2}\mathbf{u}_i\otimes\mathbf{u}_i-\frac{1}{2}\mathbb{I}\bigg)
\end{equation} where $\mathbf{u}_i$ is the pseudo-nematic director pointing along the long axis of the $i$th rod, $\otimes$ denotes the outer product and $\mathbb{I}$ is the identity matrix. Similarly, the helical order is defined as the largest eigenvalue of the tensor,
\begin{equation}
\mathbb{M}=\frac{1}{N}\sum_i^N(\mathbf{u}_i\otimes\mathbf{m}_i+\mathbf{m}_i\otimes\mathbf{u}_i)
\end{equation}
where $\mathbf{m}_i=\mathbf{u}_i\times\mathbf{b}_i$ is the out-of-plane director of the $i$th rod. The smectic order parameter is defined as
\begin{equation}
\langle \rho \rangle=\frac{1}{N}\sum_j^N e^{\frac{2\pi i}{d}\mathbf{n}\cdot\mathbf{r}_j}
\end{equation}
where $d$ is the optimal layer spacing that maximizes $\langle \rho \rangle$, $\mathbf{n}$ is the nematic director and $\mathbf{r_j}$ is the midpoint of the $j$th rod's backbone. Here, we take $\mathbf{n}$ to be the eigenvector associated with the largest eigenvalue of $\mathbb{Q}$. Figure~\ref{SI: fluid OP} shows these order parameters calculated for the example of shear oriented parallel to the helical axis of the nematic twist-bend phase. Previous numerical studies of the nematic twist-bend phase of bent-core liquid crystals report values of $\langle S \rangle\in[0.25,0.35]$, $\langle\vert\tau\vert\rangle\in[0.55,0.85]$ and $\langle \rho\rangle\lesssim0.05$. The results reported in here and in Figure~\ref{fig: solid like orientation}b are in good agreement with the values at the beginning and end of each shear simulation, allowing us to identify these states as nematic twist-bend.

\section{Coarse Graining Bend and Nematic Directors}\label{cg appendix}

To analyze the topological defects in a given configuration, we characterize the orientation of every rod and coarse grain this information to obtain a lattice of average bend and nematic directors. As the curvature breaks the azimuthal symmetry of the rods, an orthonormal frame $\{\mathbf{u},\mathbf{b},\mathbf{m}\}$ is needed to describe their biaxial orientation. Here, $\mathbf{u}$ denotes the pseudo-nematic director corresponding with the rods' long axis, $\mathbf{b}$ defines the bend director aligned with the radius of curvature and $\mathbf{m}=\mathbf{u}\times\mathbf{b}$ is the out-of-plane director. These vectors are easily defined in terms of the sequentially ordered positions, $\mathbf{r}_i$ of beads making up the simulated rods, with $i\in[0,15]$. For instance, the out-of-plane director can be defined as:
\begin{equation}
\mathbf{m}=\frac{\mathbf{v}_1\times\mathbf{v}_2}{\vert \mathbf{v}_1\times\mathbf{v}_2\vert}
\end{equation}
where $\mathbf{v}_1=\mathbf{r}_{15}-\mathbf{r}_{\rm c}$, $\mathbf{v}_2=\mathbf{r}_{0}-\mathbf{r}_{\rm c}$ and $\mathbf{r}_{\rm c}=(\mathbf{r}_8+\mathbf{r}_9)/2$. Similarly, the bend vector can be defined as:
\begin{equation}
\mathbf{b}=\frac{\mathbf{r}_{\rm b}-\mathbf{r}_{\rm c}}{\vert \mathbf{r}_{\rm b}-\mathbf{r}_{\rm c}\vert}
\end{equation}
where $\mathbf{r}_{\rm c}(\mathbf{r}_{15}+\mathbf{r}_0)/2$. The pseudo-nematic vector can be expressed as:
\begin{equation}
\mathbf{u}=\frac{\mathbf{m}\times\mathbf{u}}{\vert\mathbf{m}\times\mathbf{u}\vert}.
\end{equation}
In  addition, we can also define a local pseudo-nematic director:
\begin{equation}
\mathbf{u}_i=\frac{\mathbf{r}_{i+1}-\mathbf{r}_i}{\vert \mathbf{r}_{i+1}-\mathbf{r}_i\vert}
\end{equation}
and local bend:
\begin{equation}
\mathbf{b}_i=\mathbf{m}\times\mathbf{u}_i
\end{equation}
aligned along the bond vector between nearest neighbor beads. Next, we partition the simulation box into a $100\times100\times100$ grid of voxels with side length $\sim1\sigma$. The center of each voxel represents a site on the coarse grained lattice. We then identify a list of every $\mathbf{b}_i$ and $\mathbf{u}_i$ emanating from a bead inside the first and second nearest neighbor voxel shell around each site. From this, a coarse-grained lattice of bend directors is obtained by averaging all of the $\mathbf{b}_i$'s associated to each site. To deal with the nematic symmetry of the local pseudo-nematic director, we average the local Q-tensor:
\begin{equation}
q_i=\frac{3}{2}(\mathbf{u}_i\otimes\mathbf{u}_i-\frac{1}{3}\mathbb{I})
\end{equation}
at every site. Here $\mathbb{I}$ is the identity matrix. With this, the coarse-grained lattice of nematic directors is obtained by finding the eigenvector associated with the largest eigenvalue of the average Q-tensor at each site. This procedure was carried out for every frame of our shear simulation, allowing the evolution of defects to be tracked as a function of strain. The local $\mathbf{u}_i$ and $\mathbf{b_i}$ were used in conjunction with the overlapping voxel shells, as we found this allows a finer coarse-grained mesh and smoother gradients in bend and nematic director.

\section{Identifying Bend Defects}\label{bend appendix}

In bent-core liquid crystals, line-like geometric singularities of bend correspond to topological defects~\cite{binysh2020geometry}. Using the coarse grained lattice of bend directors, these structures are easily identified as connected clusters of lattice sites with $\vert\mathbf{b}\vert\lesssim0.1$. Moreover, the defect density (as plotted in Figure~\ref{fig: defects}a) is defined as the fraction of lattice sites belonging to one of these clusters.

\section{Measuring Topological Charge}\label{charge appendix}

\begin{figure}[ht!]%
\includegraphics[width=0.45\textwidth]{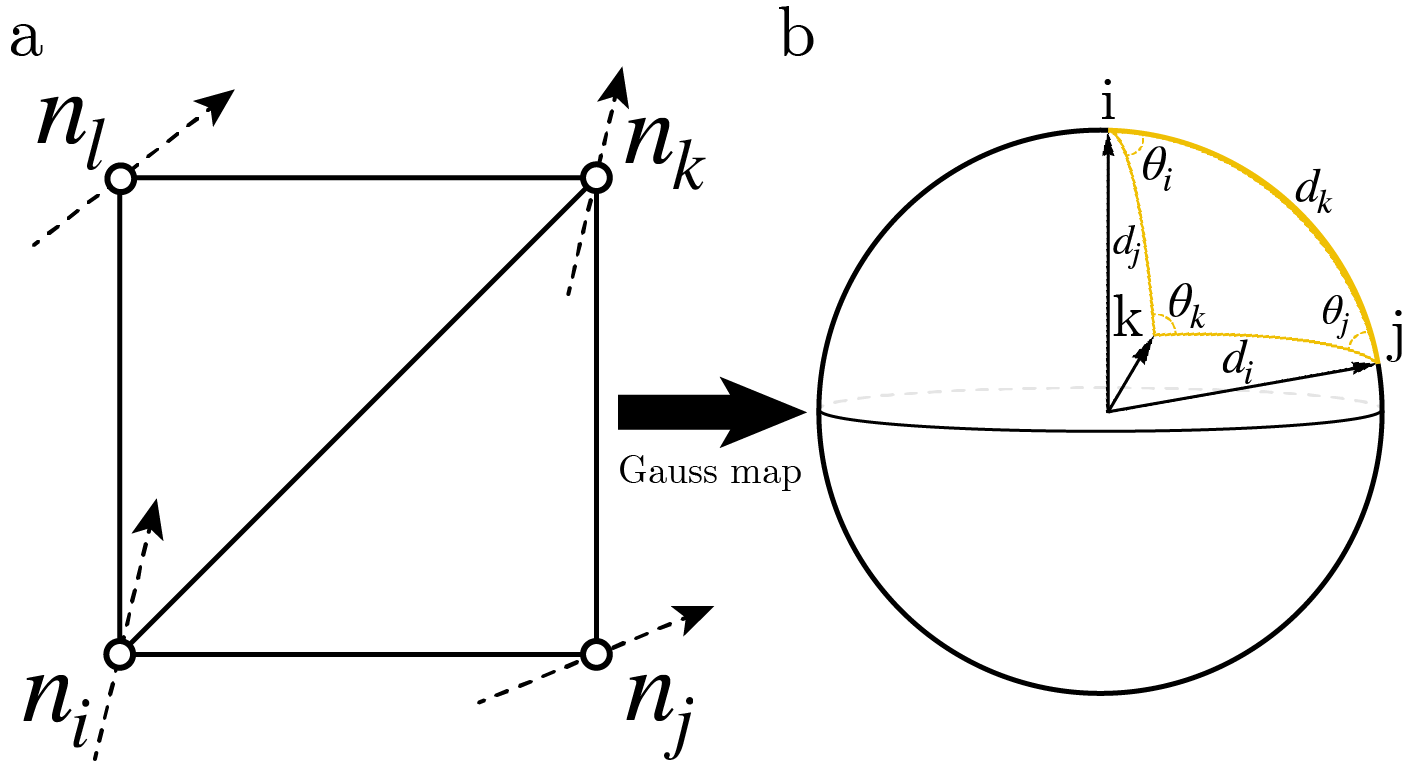}
\caption{\textbf{Schematic Illustration of Gauss map of lattice plaquette triangles onto surface of unit sphere} \textbf{a} Coarse grained lattice plaquette of nematic directors $\mathbf{n}_i, \mathbf{n}_j,\mathbf{n}_k$ and $\mathbf{n}_l$  divided into two triangular cells. \textbf{b} Image of Gaussian map of directors $\mathbf{n}_i, \mathbf{n}_j,\mathbf{n}_k$ onto unit sphere. Gold lines depicts edges of a spherical triangle whose area contributes to the discrete topological charge sum. Annotated interior angles and edges illustrated here match the definitions provided in text.}   \label{fig: gauss map}
\end{figure}

Non-singular defects in systems with a 3D order parameter are characterized by their topological charge, Q, which is often expressed in integral form as:
\begin{equation}
Q=\frac{1}{4\pi}\int \rm dx\rm dy ~\mathbf{n}\cdot\partial_x\mathbf{n}\times\partial_y\mathbf{n}.
\end{equation}
While this form is simple to write down, it can be difficult to apply to the result of numerical simulations. Thus, it is helpful to realize that $Q$ has a simple interpretation as the surface area fraction of the unit sphere which is covered by the Gaussian map of $\mathbf{n}$ over some 2D measuring surface. Using this, it is intuitive to define a discrete topological charge~\cite{berg1981definition},
\begin{equation}
Q=\frac{1}{4\pi}\sum_{\langle ijk\rangle}(\sigma A)(\mathbf{n}_i,\mathbf{n}_j,\mathbf{n}_k)
\end{equation}\label{eq: discrete charge}
which can easily be evaluated as the sum over half-plaquette triangles $\langle ijk\rangle$ of the coarse-grained lattice of nematic directors. Here,
\begin{equation}
\sigma(\mathbf{n}_i,\mathbf{n}_j,\mathbf{n}_k)=\rm sgn(\mathbf{n}_i\cdot\mathbf{n}_j\times\mathbf{n}_k)
\end{equation}
defines the sign of the oriented area of the spherical triangle that has vertices located at the image of $\mathbf{n}_i,\mathbf{n}_j,\mathbf{n}_k$ under the Gaussian map. Specifically, this area can be defined as:
\begin{equation}
A=\theta_i+\theta_j+\theta_k-\pi.
\end{equation}
Here $\theta_i$ is the interior angle at vertex $i$ of the spherical triangle and can be obtained from the spherical law of cosines via:
\begin{equation}
\cos\theta_i=\frac{\cos d_i-\cos d_j\cos d_k}{\sin d_j\sin d_k}
\end{equation}
where
\begin{equation}
d_i=\cos^{-1}(\mathbf{n}_j\cdot\mathbf{n}_k)
\end{equation}
is the arc-length of the edge opposite from angle $\theta_i$. Similar expressions for the other interior angles can be obtained by permuting $ijk$ in the above expressions. A schematic illustration of a half-plaquette triangle and its corresponding Gaussian map image is given in Figure~\ref{fig: gauss map}.

It is relevant to point out that a direct application of equation~\ref{eq: discrete charge}1 to our course grained lattice is complicated by the nematic symmetry of the director field. This is because flipping the sign of $\mathbf{n}$ on each vertex of a triangle yields eight configurations that are equivalent in real space but whose image under the Gaussian map can have different area contributions. Assuming smooth variations of $\mathbf{n}$ within a single plaquette, it is reasonable to take the configuration with the smallest area contribution to be the most topologically relevant. However, there is an additional complication that configurations with the same area but opposite orientations (i.e. $\sigma=\pm 1$) will exist. Thus, we further require plaquette triangles to be consistently oriented with respect to their neighbors. Both of these issues are largely resolved by applying the topology accommodating direction assignment algorithm to the lattice plane before calculating the charge~\cite{saha2022tada}. This coherently aligns $\mathbf{n}$ everywhere except for plaquettes that either contain a disclination or traverse a branch cut in orientation emanating from one. In the case of plaquettes crossing a branch cut, the issue can be manually resolved by finding the minimal number of directors that need to be flipped to ensure the sign of the dot product along every edge of the  plaquette\textemdash as well as the diagonal edge shared between the two triangles\textemdash is positive. Conversely, plaquettes containing disclinations cannot be consistently oriented along every edge and will therefore contain two triangles with opposite orientation. As the area contribution of these two triangles will largely cancel out, we omit them from the sum to avoid having to choose a random orientation. We further justify this omission by noting that the core region of both merons and Skyrmions is expected to be non-singular so the disclination points generically lie outside the region of double twist and do not effect the final charge. In summary, the cleaning procedure outlined here is implemented to ensure the image of a given lattice configuration yields a smooth triangulation of the unit sphere, like the example shown in Figure~\ref{fig: defects}d.

\begin{figure*}[ht!]%
\includegraphics[width=\textwidth]{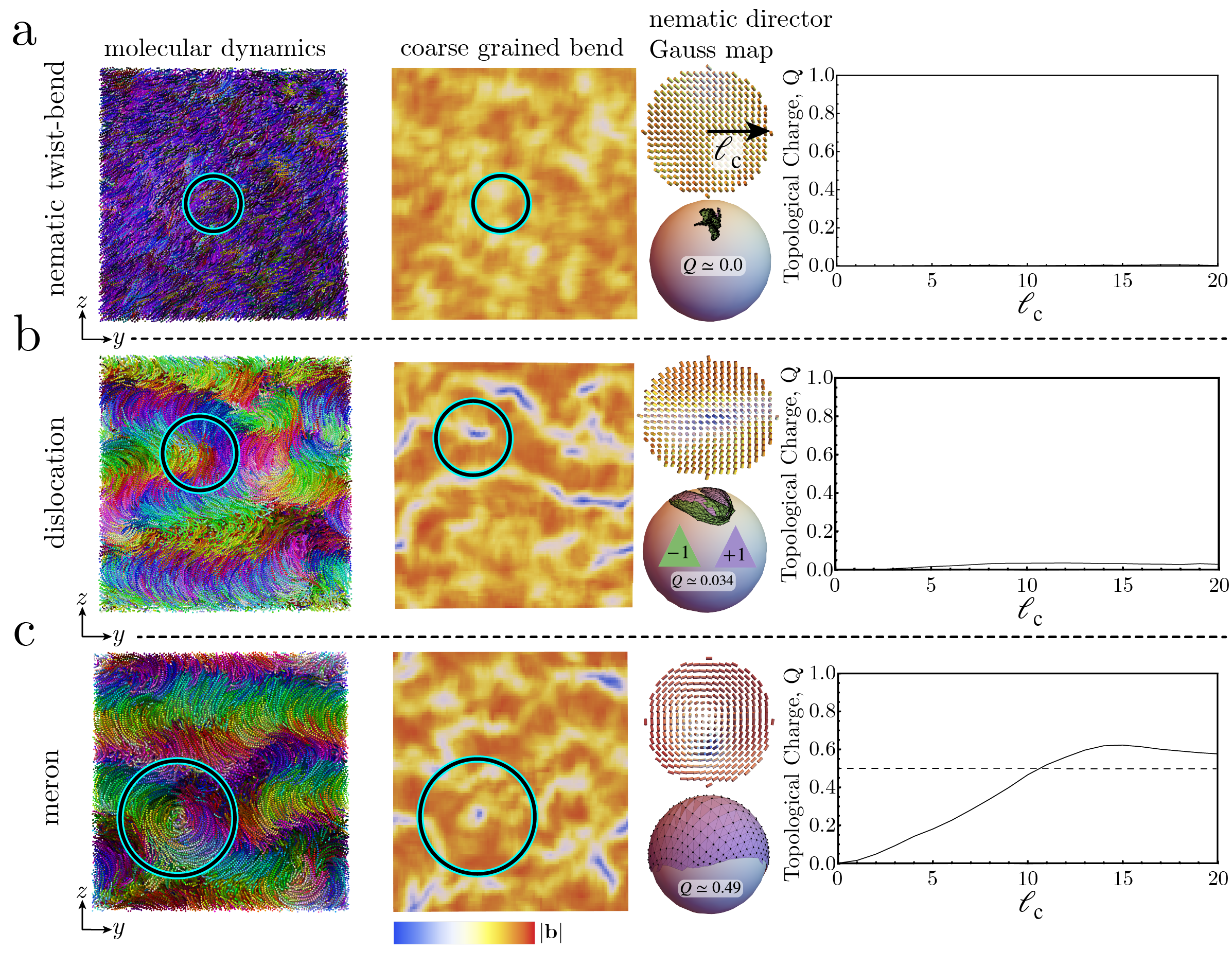}
\caption{ \textbf{Analysis of topological charge around bend zeros.} Slice of molecular dynamics data and corresponding coarse grained bend lattice shown for \textbf{a} defect free nematic twist-bend state, \textbf{b} screw/edge dislocation with no topological (i.e. Skyrmion) charge and \textbf{c} fractionally charged meron defect. Black and blue circles encapsulate location of bend zero being analyzed. Coarse grained nematic texture and image of Gaussian map shown for disc of radius $\ell_{\rm c}=10$ around the bend zero core for \textbf{b}\&\textbf{c}. Nematic director and Gaussian map image for defect free case (\textbf{a}) shown on disc of radius $\ell_{\rm c}=10$ around center of the plane. Color of triangular plaquette in Gaussian map image denotes sign $\sigma$ of the oriented area, with $\sigma=+1$ represented as purple triangles and $\sigma=+1$ represented as green ones. Topological charge as function of measuring surface radius $\ell_{\rm c}$ provided for each example.  } \label{SI: defect charge}
\end{figure*}

Evaluating the sum in eq.~\ref{eq: discrete charge}1 over every plaquette in a lattice plane gives the total topological charge of all the defects whose core intersects the plane. Thus, to characterize the charge distribution of a single bend degeneracy, it is necessary to restrict ourselves to measuring surfaces that are only intersected by one defect line. However, the topological charge is not confined to discrete values when evaluated over open measuring surfaces. For example, it is easy to see that the image of the Gaussian map of coarse grained $\mathbf{n}$ around the core of a Skyrmion would not cover the unit sphere. In fact, the measured charge will depend on the size of the measuring surface. Fortunately, the double twist defects observed here have a finite size that roughly scales with the radius of curvature of the constituent rods. In addition, the defects observed here are embedded in a charge neutral nematic twist-bend texture. Thus, we expect the dependence of $Q$ on the area of the measuring surface will only persist up to a length scale commensurate with the defect size.

We therefore characterize the charge of a particular defect line using the following procedure. We identify a perpendicular lattice plane bisecting the bend zero of interest and orient the coarse grained directors. We then calculate $Q$ over square patches with side length $2\ell_{\rm c}$, centered around the defect line, and vary $\ell_{\rm c}$ from $1$ to $20$. Examples of $Q$ as a function of $\ell_{\rm c}$ are shown in the rightmost column of Figure~\ref{SI: defect charge} for three different bend degeneracies. From here, we see that the charge around one of these bend zeros gradually increases with $\ell_{\rm c}$ before plateauing to a value of $Q\sim 0.5$. In contrast, the charge around the other two bend zero does not depend strongly on size and remain below $Q<0.05$ for all $\ell_{\rm c}$. Using this, we categorize bend zeros as dislocations if their charge plateaus at $Q\lesssim0.35$, merons if $Q\in[0.35,0.65]$ and Skyrmions if $Q\sim1$. However, as we expect that the merons have a core size controlled by the radius of curvature of the rods, all specifically reported defect charges correspond to charge evaluated over a measuring surface of size $\ell_{\rm c}=10\sigma.$ Also, it is relevant to note that no bend zeros with $Q\sim1$ were observed in any of our shear simulations, leading us to conclude that only merons are nucleated under shear. Detailed examples of the coarse grained director lattice, Gaussian map and $Q$ vs. $\ell_{\rm c}$ plot for defect free twist-bend state, screw/edge dislocation and meron defect are provided in Figure~\ref{SI: defect charge}.

\bibliography{apssamp}

\end{document}